\documentclass[11pt]{iopart}
\usepackage{bm}
\usepackage{hyperref}
\usepackage{amssymb}
\usepackage{amsopn}
\usepackage{graphicx}
\renewcommand{\etal}{\emph{et al.}}

\newcommand{\fnl}{f_{\mathrm{NL}}}
\newcommand{\fnla}{f_{\mathrm{NL1}}}
\newcommand{\fnlb}{f_{\mathrm{NL2}}}
\newcommand{\gnl}{g_{\mathrm{NL}}}

\newcommand{\Mp}{M_{\mathrm{P}}}

\renewcommand{\d}{\mathrm{d}}
\newcommand{\vect}[1]{\bm{\mathrm{{#1}}}}

\renewcommand{\geq}{\geqslant}

\makeatletter
\newcommand\numberwithin[2]{\@addtoreset{#1}{#2}}
\makeatother
\numberwithin{footnote}{section}

\begin{document}
	\title{Moment transport equations for non-Gaussianity}
%	\date{\today}
	\author{David J. Mulryne$^1$, David Seery$^1$ and Daniel Wesley$^{1,2}$}
	\address{$^1$ Department of Applied Mathematics and Theoretical Physics \\
	Wilberforce Road, Cambridge, CB3 0WA, United Kingdom \\
	$^2$ Center for Particle Cosmology\\
	David Rittenhouse Laboratory, University of Pennsylvania \\
	209 South 33rd Street, Philadelphia, PA 19104 USA
	}
	\eads{\mailto{djm63@cam.ac.uk}, \mailto{djs61@cam.ac.uk},
		\mailto{dwes@sas.upenn.edu}}
	\submitto{JCAP}
	\begin{abstract}
		We present a novel method for calculating the primordial
		non-Gaussianity produced by super-horizon evolution during inflation.
		Our method evolves
		the distribution of coarse-grained inflationary field
		values using a transport equation.
		We present simple evolution equations for the moments of this
		distribution, such as the variance and skewness.
		This method possesses
		some
		advantages over existing techniques. Among them, it
		cleanly separates multiple sources of primordial non-Gaussianity,
		and
		is computationally efficient
		when compared with popular alternatives,
		such as the $\delta N$ framework.
		We adduce numerical calculations demonstrating that our new method
		offers good agreement with those already in the literature.
		We focus on two fields and the $\fnl$ parameter,
		but we expect our method will
		generalize to multiple scalar fields and to moments of
		arbitrarily high order.
		We present our expressions in a
		field-space covariant form which we postulate to be valid for any
		number of fields.
		
	\vspace{3mm}
	\begin{flushleft}
	\end{flushleft}
	\end{abstract}
	\maketitle
	
	\section{Introduction}
		
	Inflation generically predicts a primordial spectrum of density
	perturbations which is almost precisely Gaussian
	\cite{Hawking:1982cz,Hawking:1982my,
	Bardeen:1983qw,Lyth:1984gv,Guth:1985ya}.
	In recent years the small non-Gaussian component 
	\cite{Falk:1992sf,Gangui:1993tt,Pyne:1995bs,
	Acquaviva:2002ud,Maldacena:2002vr,
	Lyth:2005fi,Seery:2005gb}
	has emerged as an
	important observable
	\cite{Komatsu:2009kd}, and will be measured
	with good precision by the
	\emph{Planck Surveyor} satellite
	\cite{Komatsu:2001rj}.
	In the near future, as observational data become more plentiful, it
	will be important to understand the non-Gaussian
	signal expected in a wide variety of
	models, and to anticipate what conclusions can be drawn about
	early-universe physics from a prospective detection of primordial
	non-Gaussianity.
	
	In this paper, we present a novel method
	for calculating the primordial
	non-Gaussianity produced by super-horizon evolution in two-field models
	of inflation.	Our method is based on the real-space
	distribution of inflationary field values on a flat hypersurface,
	which can be thought of as a probability density function
	whose evolution is determined by a form of the collisionless
	Boltzmann equation. Using a cumulant representation
	\cite{Juszkiewicz:1993hm,Bouchet:1993xb,Bouchet:1995ez,Fosalba:1999ha}
	to expand our density function
	around an exact Gaussian,
	we derive ordinary differential equations
	which evolve the moments of this distribution.
	Further, we show how these moments are related to observable quantities,
	such as the dimensionless bispectrum measured by $\fnl$
	\cite{Komatsu:2001rj,Maldacena:2002vr}.
	We present numerical
	results which show that this method gives good agreement with other
	techniques. It is not necessary
	to make any assumptions about the inflationary model beyond
	requiring a canonical kinetic term and applying the slow-roll
	approximation. While there are already numerous methods for computing the
	super-horizon contribution to $\fnl$, including the widely used
	$\delta N$ formalism, we believe the one reported here has a number of
	advantages.  
	
	First, this new technique is ideally suited to unraveling the various
	contributions to $\fnl$.  
	This is because we follow the moments of the inflaton
	distribution directly, which makes it straightforward
	to identify large contributions to the skewness or other moments.
	The evolution equation for each moment is simple and possesses clearly
	identifiable source terms, which are related to the properties of the
	inflationary flow on field space.
	This offers a clear separation between two key sources of primordial
	non-Gaussianity. One of these is the intrinsic non-linearity
	associated with evolution of the
	probability density function
	between successive flat hypersurfaces;
	the other is a gauge transformation from field fluctuations to
	the  curvature peturbation, $\zeta$.
	It would be difficult or impossible to observe this split
	within the context of other calculational schemes, such as the
	conventional $\delta N$ formalism.

	A second advantage of our method is connected with the computational
	cost of numerical implementation.
	Analytic formulas for $\fnl$ are known in certain cases, mostly
	in the context of the $\delta N$ framework, but only for very
	specific choices of the potential
	\cite{Vernizzi:2006ve,Battefeld:2006sz,Byrnes:2008wi,Byrnes:2008zy}
	or Hubble rate \cite{Byrnes:2009qy,Battefeld:2009ym}.
	These formulas become increasingly cumbersome as the number of fields
	increases, or if one studies higher moments
	\cite{Seery:2006js,Byrnes:2006vq}.
	In the future, it seems clear that
	studies of complex models with many fields will
	increasingly rely on numerical methods.
	The numerical
	$\delta N$ framework requires the solution to a number of
	ordinary differential equations which scales exponentially with the
	number of fields.
	Since some models include hundreds of fields,
	this may present a significant obstacle \cite{Dimopoulos:2005ac}.
	Moreover,
	the $\delta N$ formalism depends crucially on a
	numerical integration algorithm with low noise properties,
	since finite differences must be extracted between perturbatively
	different initial conditions after $\sim 60$ e-folds of evolution.
	Thus, the background equations must be solved to great accuracy, 
	since any small error has considerable scope to propagate.	
	
	In this paper we ultimately solve our equations numerically
	to determine the evolution of moments in specific models.
	Our method requires the solution to a number of differential
	equations which scales at most polynomially (or in certain cases perhaps
	even linearly) with the number of fields.
	It does not rely on extracting finite differences, and therefore is
	much less susceptible to numerical noise.
	These advantages may be
	shared with other schemes, such as the numerical
	method recently employed by Lehners \& Renaux-Petel
	\cite{Lehners:2009ja}.

	A third advantage, to which we hope to return in a future publication,
	is that our formalism yields explicit evolution equations
	with source terms.
	From an analysis of these source terms, we hope that it will
	be possible to identify those physical features
	of specific models
	which lead to the production
	of large non-Gaussianities.

	This paper is organized as
	follows.  
	In \S\ref{sec:computing_fnl}, we show how the
	non-Gaussian parameter $\fnl$ can be computed in our framework. 
	The calculation remains in real space throughout
	(as opposed to Fourier space), which modifies
	the relationship between $\fnl$ and the multi-point functions of the
	inflaton field.
	Our expression for $\fnl$ shows a
	clean separation between different contributions to non-Gaussianity,
	especially between the intrinsic nonlinearity of the field evolution and
	the gauge transformation between comoving and flat hypersurfaces.	
	In \S\ref{sec:transport}, we introduce our model for the
	distribution of inflaton field values, which is a ``moment expansion"
	around a purely Gaussian distribution.	 We derive the equations which
	govern the evolution of the moments of this distribution in the one- and
	two-field cases. In
	\S\ref{sec:numerics}, 
	we present a comparison of our new technique and those already in the
	literature. We compute $\fnl$ numerically in several two-field models,
	and find excellent agreement between techniques.
	We conclude in \S\ref{s:conclusions}.
	
	Throughout this paper, we use units in which $c = \hbar = 1$,
	and the reduced Planck mass
	$\Mp^{-2} \equiv 8 \pi G$ is set to unity.

	\section{Frameworks for computing $\fnl$}
	\label{sec:computing_fnl}
	
	In this section, we introduce our new method for computing the
	non-Gaussianity parameter $\fnl$.	This method requires three main
	ingredients: a formula for the curvature perturbation, $\zeta$, in
	terms of the field values on a spatially flat hypersurface;
	expressions for the derivatives of the number of e-foldings, $N$,
	as a function of field values at horizon exit;
	and a prescription for evolving the field distribution from horizon exit
	to the time when we require the statistical properties of $\zeta$.
	The first two ingredients are given in
	Eqs.~\eref{e:correlators2}--\eref{e:correlators3}
	and \eref{e:Nphi}--\eref{e:Nphichi},
	found at the end of \S\ref{ss:sep_universe} and
	\S\ref{sec:derivative-N} respectively.
	The final ingredient is discussed in \S\ref{sec:transport}.
	
	\subsection{Calculations beyond linear order}
	\label{sec:overview}
	
	Once it became clear that non-linearities of the microwave
	background anisotropies could be detected
	by the WMAP and \emph{Planck} survey
	satellites \cite{Komatsu:2001rj},
	many authors studied
	higher-order correlations of the curvature perturbation.
	In early work,
	direct calculations of a correlation function
	were matched to the known limit of local non-gaussianity
	\cite{Acquaviva:2002ud,
	Maldacena:2002vr,Creminelli:2003iq,Zaldarriaga:2003my,
	ArkaniHamed:2003uz,Alishahiha:2004eh}. This method works well
	if isocurvature modes are absent, so that the curvature perturbation
	is constant after horizon exit.
	In the more realistic situation that isocurvature modes cause
	evolution on superhorizon scales, all correlation functions become
	time dependent.
	Various formalisms have been employed
	to describe this evolution.
	Lyth \& Rodr\'{\i}guez \cite{Lyth:2005fi}
	extended the $\delta N$ method
	\cite{Starobinsky:1986fxa,Sasaki:1995aw}
	beyond linear order.
	This method is simple and well-suited to analytical
	calculation.
	Rigopoulos, Shellard and van Tent
	\cite{Rigopoulos:2004gr,Rigopoulos:2005xx}
	worked with a gradient expansion, rewriting the field equations in
	Langevin form. The noise term
	was used as a proxy for setting initial
	conditions at horizon crossing.
	A similar `exact' gradient formalism was written down by
	Langlois \& Vernizzi
	\cite{Langlois:2005ii,Langlois:2005qp,Langlois:2006vv}.
	In its perturbative form, this approach has been used by
	Lehners \& Renaux-Petel to obtain numerical results
	\cite{Lehners:2009ja}.
	Another numerical scheme has been introduced by
	Huston \& Malik \cite{Huston:2009ac}.

	What properties do we require of a successful prediction?
	Consider a typical observer, drawn at random from an ensemble of
	realizations of inflation.
	In any of the formalisms discussed above,
	we aim to estimate the statistical
	properties of the curvature perturbation which would be measured by
	such an observer.
	Some realizations
	may yield statistical properties which are quite
	different from the ensemble average, but
	these
	large excursions are uninteresting
	unless anthropic arguments are in play.

	Next we introduce a collection of
	comparably-sized spacetime volumes %(``separate universes''),
	whose mutual scatter is destined
	to dominate the microwave background anisotropy on a given
	scale. Neglecting spatial gradients,
	each spacetime volume will follow a trajectory in
	field space which is slightly displaced from its neighbors.
	The scatter between trajectories is determined by initial conditions
	set at horizon exit, which
    are determined
	by promoting the vacuum
	fluctuation to a classical perturbation.%
		\footnote{The scatter among
		trajectories may grow with the overall
		volume, owing to back-reaction effects in de Sitter space
		\cite{Lyth:2007jh,Sloth:2006az,Sloth:2006nu,Seery:2007wf}.
		If inflation can end in more than one
		vacuum, leading to
		different predictions for observable quantities, it may be
		helpful to evaluate this enhanced dispersion for the purpose
		of deciding into which vacuum the field will fall.
		Once this minimum has been decided, however, large-volume effects
		needlessly complicate the calculation. In this paper,
		we will always assume that predictions are being made by
		following a sufficient number of trajectories to determine
		the statistical properties with reasonable precision, but no more.}
	A correct prediction is a function of the trajectories followed by
	every volume in the collection, taken as a whole.
	One never makes a prediction for a single trajectory.
	
	Each spacetime volume follows a trajectory, which we label with
	its position $\varphi^\star$ at some fixed time, to be made
	precise below.
	Throughout this paper, superscript `$\star$' denotes evaluation on
	a spatially flat hypersurface.
	Consider the evolution of some quantity of interest, $F$, which is
	a function of trajectory.
	If we know the distribution $P(\varphi^\star)$
	we can study statistical properties of $F$
	such as the $m^{\mathrm{th}}$ moment $\kappa_m$,
	\begin{equation}
		\label{eq:separate-universe-moment}
		\kappa_m
		\equiv
		\int \d\varphi^\star \; P(\varphi^\star)
		\left[
			F(\varphi^\star) - \langle F \rangle
		\right]^m ,
	\end{equation}
	where we have introduced the ensemble average of $F$,
	\begin{equation}
		\label{eq:ensemble-average}
		\langle F \rangle
		\equiv
		\int \d\varphi^\star \; P(\varphi^\star)
		F(\varphi^\star) .
	\end{equation}
	In Eqs.~\eref{eq:separate-universe-moment}--%
	\eref{eq:ensemble-average},
	$\varphi^\star$ stands for a collection
	of any number of fields.
	It is the $\kappa_m$ which are observable quantities.

	Eq.~\eref{eq:separate-universe-moment} defines what
	we will call the
	exact separate universe picture.
	It is often convenient
	to expand $F(\varphi^\star)$
	as a power series in the field values
	centered on a fiducial trajectory,
	labelled `fid,'
	\begin{equation}
		\label{eq:perturbative-deltan}
		F(\varphi^\star) - F(\varphi^\star_{\mathrm{fid}}) =
		\sum_{n=1}^{\infty}
		\frac{1}{n!}
		( \varphi^\star - \varphi^\star_{\mathrm{fid}} )^n
		\left.
		\frac{\partial^n F}{\partial (\varphi^\star)^n}
		\right|_{\varphi^\star = \varphi^\star_{\mathrm{fid}}}
		.
	\end{equation}
	When Eq.~\eref{eq:perturbative-deltan} is used to evaluate
	the $\kappa_m$, we refer to
	the `perturbative' separate universe picture.
    If all terms in the power series are retained, these two versions
	of the calculation are formally equivalent.
	In unfavorable cases, however,
	convergence may occur slowly or not at all.
	This possibility was discussed in
	Refs.~\cite{Cogollo:2008bi,Rodriguez:2008hy}.
	Although our calculation is formally perturbative, it is not
	directly equivalent to Eq.~\eref{eq:perturbative-deltan}.
	We briefly discuss the relation of our calculation to conventional
	perturbation theory in \S\ref{s:conclusions}.
		
	\subsection{Calculating $\fnl$ in the separate universe picture}
	\label{ss:sep_universe}
	
	By definition, the curvature perturbation $\zeta$
	measures local fluctuations in expansion history
	(expressed in e-folds $N$),
	calculated on a comoving hypersurface.
	In many models, the curvature perturbation
	is synthesized by
	superhorizon physics, which reprocesses a
	set of Gaussian fluctuations generated at horizon exit.
	In a single-field model, only one Gaussian fluctuation
	can be present, which we label $\zeta_g$.
	Neglecting spatial gradients,
	the total curvature perturbation must
	then be a function of $\zeta_g$ alone.
	For $|\zeta_g| \ll 1$, this can be
	well-approximated by
	\begin{equation}
		\zeta \simeq \zeta_g + \frac{3}{5} \fnl \left(
			\zeta_g^2 - \langle \zeta_g^2 \rangle
		\right) ,
		\label{eq:local-ng}
	\end{equation}
	where $\fnl$ is independent of spatial position.
	Eq.~\eref{eq:local-ng} defines the so-called ``local'' form of
	non-gaussianity. It applies only when quantum interference effects
	can be neglected, making $\zeta$ a well-defined object rather than a
	superposition of operators
	\cite{Seery:2008ax}. If this condition is satisfied,
	spatial correlations of $\zeta$ may be extracted and it
	follows that $\fnl$ can be estimated using the rule
	\begin{equation}
		\label{eq:fnl-basic}
		\fnl \simeq
			\frac{5}{18}
			\frac{\langle \zeta \zeta \zeta \rangle}
				{\langle \zeta \zeta \rangle^2} ,
	\end{equation}
	where we have recalled
	that $\zeta$ is nearly Gaussian,
	or equivalently that $|\fnl| \ll |\zeta_g|^{-1}$.

	With $\fnl$ spatially independent,
	Eq.~\eref{eq:local-ng} strictly applies only in single-field inflation.
	In this case one can accurately determine $\fnl$ by applying
	Eq.~\eref{eq:local-ng} to a single trajectory
	with fixed initial conditions, as in the method
	of Lehners \& Renaux-Petel \cite{Lehners:2009ja}.
	Where more than one field is present, $\fnl$ may vary in space
	because it depends on the isocurvature modes. In this case one must
	determine $\fnl$ statistically on a bundle of adjacent
	trajectories which sample the local distribution of
	isocurvature modes.
	Eq.~\eref{eq:fnl-basic} is then indispensible.
	Following Maldacena \cite{Maldacena:2002vr},
	and later Lyth \& Rodr\'{\i}guez \cite{Lyth:2005fi},
	we adopt Eq.~\eref{eq:fnl-basic} as our definition of $\fnl$,
	whatever its origin.
	In real space, the coefficient $5/18$
	in Eq.~\eref{eq:fnl-basic}
	depends on the convention
	$\langle \zeta \rangle = 0$. %in Eq.~\eref{eq:local-ng}.
	More generally, this follows from the
	definition of $\kappa_m$, Eq.~\eref{eq:separate-universe-moment}.
	In Fourier space, either prescription is automatically enforced
	after dropping disconnected contributions, again leading to
	Eq.~\eref{eq:fnl-basic}.

	To proceed, we require estimates of the correlation functions
	$\langle \zeta \zeta \rangle$ and
	$\langle \zeta \zeta \zeta \rangle$.
	We first describe the conventional approach, in which `$\star$'
	denotes a flat
	hypersurface at a fixed initial time.	
	The quantity $N(\varphi^\star_{i})$ denotes the number of e-foldings
	between this initial slice and a final comoving hypersurface,
	where $i$ indexes the species of light scalar fields.
	The local variation in expansion can be written in the fiducial picture
	as
	\begin{equation}
		\label{eq:deltaN}
		\zeta(\vect{x}) \equiv \delta N(\vect{x})
		= \sum_{n=1}^{\infty}
		\frac{1}{n!}
		\left.
		\frac{\partial^n N(\varphi^\star_i)}
			{\partial \varphi_{j_1}^\star \cdots \partial \varphi_{j_n}^\star}
		\right|_{\varphi^\star_i = \varphi^\star_{i,\mathrm{fid}}}
		\delta\varphi^\star_{j_1}(\vect{x}) \cdots
		\delta\varphi^\star_{j_n}(\vect{x}) 
		,
	\end{equation}
	where
	$\delta\varphi^\star_{j} \equiv
	\varphi^\star_{j} - \varphi^\star_{j,\mathrm{fid}}$.
	
	Subject to the condition that the relevant scales are all
	outside the horizon,
	we are free to choose the initial time---set by the hypersurface
	`$\star$'---at our convenience.
	In the conventional approach,
	`$\star$' is taken to lie a few e-folds
	after our collection of spacetime volumes passes outside the causal
	horizon \cite{Lyth:2005fi,Seery:2005gb}.
	This choice has many virtues.
	First, we need to know statistical properties of the field fluctuations
	$\delta \varphi^\star_i$ only around the time of horizon
	crossing, where they can be computed without the appearance of
	large logarithms \cite{Seery:2007wf,Seery:2008qj}.
	Second, as a consequence of the slow-roll
	approximation, the $\delta\varphi^\star_i$ are
	uncorrelated at this time, 
	leading to algebraic simplifications.
	Finally, the $\delta N$ formula %also 
	subsumes a gauge transformation
	from the
	field variables $\delta \varphi^\star_i$ to the observational
	variable $\zeta$.
	Using Eqs.~\eref{eq:separate-universe-moment}--\eref{eq:ensemble-average},
	\eref{eq:fnl-basic} and~\eref{eq:deltaN}, one finds that
	$\fnl$ can be written to a good approximation %as
	\cite{Lyth:2005fi}
	\begin{equation}
		\label{eq:fnl-deltaN}
		\fnl \approx \frac{5}{6}
		\frac{N_{,i} N_{,j} N_{,ij}}
		{(N_{,k} N_{,k})^{2}} ,
		\hfill
		\textsf{`$\star$' at horizon crossing}
		\hspace{1cm}
	\end{equation}
	where $N_{,i} \equiv \partial N/\partial \varphi^\star_i$
	and for simplicity %of expression 
	we have
	dropped the `$\star$' which indicates time of evaluation.
	A similar definition applies for $N_{,ij}$.
	
	Eq.~\eref{eq:fnl-deltaN} is accurate up to small
	intrinsic non-Gaussianities present in the field
	fluctuations at horizon exit.
	As a means of predicting $\fnl$ it is pleasingly compact,
	and straightforward to evaluate in many models.
	Unfortunately, it also obscures %whatever 
	the physics which determines
	$|\fnl|$. For this reason
	it is hard to infer,
	from Eq.~\eref{eq:fnl-deltaN} alone,
	those classes
	of models in which $|\fnl|$ is always large or small.%
		\footnote{Even in simple models
		it can be quite subtle to determine what range of
		$|\fnl|$ is dynamically allowed.
		See, for example, Refs.~\cite{Byrnes:2008wi,Byrnes:2008zy}.}
		
	Our strategy is quite different.
	We choose `$\star$' to lie
	around the time when we require the statistical properties
	of $\zeta$.
	The role of the $\delta N$ formula, Eq.~\eref{eq:deltaN},
	is then to encode \emph{only} the gauge transformation between
	the $\delta\varphi_i^\star$ and $\zeta$.
	In \S\ref{sec:derivative-N} below, we show how the appropriate 
	gauge transformation is computed using the $\delta N$ formula.
	In the present section we restrict our attention
	to determining a formula for $\fnl$ under the assumption that the 
	distribution of field values on `$\star$' is known.
	In \S\ref{sec:transport}, we will
	supply the required prescription to evolve the distribution of field
	values between horizon exit and `$\star$'.

	Although the $\delta\varphi^\star_i$ are independent
	random variables at horizon exit,
	correlations can be induced by subsequent evolution.
	One must therefore allow for off-diagonal terms
	in the two-point function. Remembering that we are working
	with a collection of spacetime volumes in real space,
	smoothed on some characteristic scale, we write
	\begin{equation}
		\label{e:correlators1}
		\langle \delta\varphi^\star_i \delta\varphi^\star_j \rangle
		\equiv \Sigma_{ij} .
	\end{equation}
	$\Sigma_{ij}$ does not vary in space, but it may be a function of
	the scale which characterizes our ensemble of spacetime volumes.
	In all but the simplest models it will vary in time.
	It is also necessary to account for intrinsic non-linearities
	among the $\delta\varphi^\star_i$, which are small at horizon
	crossing but may grow. We define
	\begin{equation}
		\label{e:correlators2}
		\langle \delta\varphi^\star_i \delta\varphi^\star_j
		\delta\varphi^\star_k \rangle
		\equiv \alpha_{ijk} .
	\end{equation}
	Likewise, $\alpha_{ijk}$ should be regarded as a function of time
	and scale.
	The permutation symmetries of an expectation value such
	as~\eref{e:correlators2} guarantee that,
	for example,
	$\alpha_{122} = \alpha_{212} = \alpha_{221}$.%
		\footnote{We are assuming that these expectation values are
		ensemble averages over classical stochastic
		fields, and are therefore
		invariant under reordering of the fields.}
	When written explicitly,
	we place the indices of symbols such
	as $\alpha$ in numerical order.
	Neglecting a small ($\lesssim \Or(\Sigma^3)$)
	intrinsic four-point correlation,
	it follows that
	\begin{equation}
		\label{e:correlators3}
		\langle \delta\varphi^\star_i \delta\varphi^\star_j
		\delta\varphi^\star_k \delta\varphi^\star_m \rangle
		=
		\Sigma_{ij} \Sigma_{km} + \Sigma_{ik} \Sigma_{jm}
		+ \Sigma_{im} \Sigma_{jk} .
	\end{equation}
   
	Now we specialize to a two-field model, parametrized by fields
	$\varphi_1$ and $\varphi_2$.
	Using Eqs.~\eref{eq:separate-universe-moment}--\eref{eq:ensemble-average},
	\eref{eq:deltaN} and~\eref{e:correlators1},
	it follows that the two-point function of $\zeta$ satisfies
	\begin{equation}
		\label{e:zetaTwo}
		\langle \zeta \zeta \rangle =
			N_{,1}^2 \Sigma_{11}
			+
			N_{,2}^2 \Sigma_{22}
			+
			2
			N_{,1} N_{,2} \Sigma_{12}
			\hfill
			\textsf{`$\star$' arbitrary}
			\hspace{1cm}
	\end{equation}
	The three-point function can be written
	\begin{equation}
		\label{e:zetaThree}
		\langle \zeta \zeta \zeta \rangle
		=
		\langle \zeta \zeta \zeta \rangle_{1}
		+
		\langle \zeta \zeta \zeta \rangle_{2} ,
	\end{equation}
	where we have identified two separate contributions, labelled
	`1' and `2'.
	The `1' term includes all contributions
	involving \emph{intrinsic} non-linearities, those which arise
	from non-Gaussian correlations among the field fluctuations,
	\begin{equation} 
		\label{e:zetathreeone}
		\fl
		\langle \zeta \zeta \zeta \rangle_{1}
		=
		N_{,1}^3 \alpha_{111}
		+
		N_{,2}^3 \alpha_{222}
		+
		3 N_{,1}^2 N_{,2} \alpha_{112}
		+
		3 N_{,1} N_{,2}^2 \alpha_{122} .
			\hfill
			\textsf{`$\star$' arbitrary}
			\hspace{1cm}
	\end{equation}
	The `2' term encodes non-linearities
	arising directly from the gauge transformation to $\zeta$
	\begin{eqnarray}
	\fl\nonumber
		\langle \zeta \zeta \zeta \rangle_{2}
		=
		\frac{9}{2} N_{,1}^2 N_{,11} \Sigma_{11}^2
		+
		\frac{9}{2} N_{,2}^2 N_{,22} \Sigma_{22}^2
			\hspace{5.3cm}
		\textsf{`$\star$' arbitrary}  \\ \mbox{} \nonumber 
					+
		9 \left(
			N_{,1} N_{,2} N_{,11} + N_{,1}^2 N_{,12}
		\right)
		\Sigma_{11} \Sigma_{12}
		+
		9 \left(
			N_{,1} N_{,2} N_{,22} + N_{,2}^2 N_{,12}
		\right)
		\Sigma_{12} \Sigma_{22}
		\\ \mbox{} \nonumber
		+
		\frac{3}{2} \left(
			N_{,2}^2 N_{,11} + N_{,1}^2 N_{,22}
			+ 4 N_{,1} N_{,2} N_{,12}
		\right)
		\left(
			\Sigma_{11} \Sigma_{22}
			+
			2 \Sigma_{12}^2
		\right)
		\\
		-
		\frac{3}{2} \left(
			N_{,11} \Sigma_{11} +
			2 N_{,12} \Sigma_{12} +
			N_{,22} \Sigma_{22}
		\right) \langle \zeta \zeta \rangle
		,
		\label{e:zetathreetwo}
	\end{eqnarray}
	After use of Eq.~\eref{eq:fnl-basic},
	Eqs.~\eref{e:zetaThree}--\eref{e:zetathreetwo} can be used to
	extract the non-linearity parameter $\fnl$. This decomposes
	likewise into two contributions $\fnl = \fnla + \fnlb$, which
	we shall discuss in more detail in \S\ref{sec:numerics}.
	
	\subsection{The derivatives of $N$}
	\label{sec:derivative-N}
	
	To compute $\fnl$ in concrete models, we require expressions for
	the derivatives $N_{,i}$ and $N_{,ij}$.
	For generic initial and final times, these are difficult to
	obtain.
	Lyth \& Rodr\'{\i}guez \cite{Lyth:2005fi} used direct integration,
	which is %successful 
	effective for quadratic potentials and constant
	slow-roll parameters.
	Vernizzi \& Wands \cite{Vernizzi:2006ve}
	obtained expressions in a two-field model with an arbitrary
	sum-separable potential by introducing
	Gaussian normal coordinates on the space of trajectories.
	Their approach was generalized to many fields by Battefeld
	\& Easther \cite{Battefeld:2006sz}.
	Product-separable potentials can be accommodated using the same
	technique \cite{Choi:2007su}.
	An alternative technique has been proposed by
	Yokoyama {\etal} \cite{Yokoyama:2007dw}.
	
	A considerable simplification occurs in the present case,
	because we only require the derivative evaluated between flat and
	comoving hypersurfaces which coincide in the unperturbed universe.
	For any species $i$, and to leading order in the slow-roll
	approximation, the number of e-folds $N$ between
	the flat hypersurface `$\star$' and a comoving hypersurface
	`$c$' satisfies
	\begin{equation}
		N \equiv
		- \int_{\varphi_i^\star}^{\varphi^c_i}
		\frac{V}{V_{,i}}
		\; \d\varphi_i
		\hfill
		\textsf{no sum on $i$,}
		\hspace{1cm}
	\end{equation}
	where $V_{,i} \equiv \partial V / \partial \varphi_i$
	and $\{ \varphi_i^\star, \varphi_i^c \}$
	are the field values evaluated on `$\star$' and `$c$,' respectively.
	Under an infinitesimal shift of $\varphi^\star_i$, we
	deduce that $N_{,i}$ obeys
	\begin{equation}
		\label{e:NfromPhiPhiStar}
		%N_{,\varphi_i^\star} =
		N_{,i} =
		\left( \frac{V}{V_{,i}} \right)^\star
		-
		\left( \frac{V}{V_{,i}} \right)^c
		\frac{\partial \varphi^c_i}{\partial \varphi^\star_i}
		\hfill
		\textsf{no sum on $i$.}
		\hspace{1cm}
	\end{equation}
	Note that this applies for an arbitrary $V$, which need not
	factorize into a sum or product of potentials for the individual
	species $i$. In principle a contribution
	from variation of the integrand is present, which spoils a na\"{\i}ve
	attempt to generalize the method of
	Refs.~\cite{Vernizzi:2006ve,Battefeld:2006sz,Choi:2007su}
	to an arbitrary potential. This contribution vanishes
	in virtue of our supposition that `$\star$' and `$c$' are
	infinitesimally separated.
	
	To compute $\partial \varphi_i^c / \partial \varphi_j^\star$
	it is helpful to introduce
	a quantity $C$,
	which in the sum-separable case coincides with the
	conserved quantity of
	Vernizzi \& Wands \cite{Vernizzi:2006ve,GarciaBellido:1995qq}.
	For our specific choice of a two-field model, this takes the
	form
	\begin{equation}
		C(\varphi_1, \varphi_2) \equiv
		\int^{\varphi_1} \frac{H^2}{V_{,1}} \; \d\varphi_1'
		-
		\int^{\varphi_2} \frac{H^2}{V_{,2}} \; \d\varphi_2' ,
	\end{equation}
	where the integrals are evaluated on a single spatial hypersurface.
	In an $M$-field model, one would obtain $M-1$ conserved quantities
	which label the isocurvature fields. The construction of these
	quantities is discussed in Refs.~\cite{Battefeld:2006sz,Seery:2006js}.
	For sum-separable potentials one can show
	using the equations of motion that $C$
	is conserved under time evolution to leading order in slow-roll.
	It is not conserved for general potentials,
	but the variation can be neglected
	for infinitesimally separated hypersurfaces.

	Under a change of trajectory, $C$ varies according to the rules
	\begin{equation}
		\label{e:C1a}
		\frac{\partial C}{\partial \varphi_1^\star} = \frac{H^2}{V_{,1}}
	\end{equation}
	and
	\begin{equation}
		\label{e:C1b}
		\frac{\partial C}{\partial \varphi_2^\star} = - \frac{H^2}{V_{,2}} .
	\end{equation}
	The comoving hypersurface `$c$' is defined by 
	\begin{equation}
		\label{e:DefOfComoving}
		\frac{1}{2} \left( \dot \varphi_1^2 + \dot \varphi_2^2 \right)
		+ V = \textrm{constant} .
	\end{equation}
	We are assuming that the slow-roll approximation applies, so that
	the kinetic energy may be neglected in comparison with the potential
	$V$. Therefore on `$c$' we have
	\begin{equation}
		\label{e:C2}
		\frac{\partial V}{\partial \varphi^c_1}
		\frac{\partial \varphi^c_1}{\partial C}
		+
		\frac{\partial V}{\partial \varphi^c_2}
		\frac{\partial \varphi^c_2}{\partial C} = 0 .
	\end{equation}
	Combining Eqs.~\eref{e:C1a}, \eref{e:C1b} and~\eref{e:C2}
	we obtain expressions for
	$\partial \varphi_i^c / \partial \varphi_j^\star$, namely
	\begin{eqnarray}
		\label{eq:1c1star}
		\frac{\partial \varphi^c_1}{\partial \varphi^\star_1}
		=
		\left( \frac{V_{,1}}{V} \right)^c
		\left( \frac{V}{V_{,1}} \right)^\star
		\sin^2 \theta , \\
		\label{eq:1c2star}
		\frac{\partial \varphi^c_1}{\partial \varphi^\star_2}
		=
		-
		\left( \frac{V_{,1}}{V} \right)^c
		\left( \frac{V}{V_{,2}} \right)^\star
		\sin^2 \theta , \\
		\label{eq:2c1star}
		\frac{\partial \varphi^c_2}{\partial \varphi^\star_1}
		=
		-
		\left( \frac{V_{,2}}{V} \right)^c
		\left( \frac{V}{V_{,1}} \right)^\star
		\cos^2 \theta , \\
		\label{eq:2c2star}
		\frac{\partial \varphi^c_2}{\partial \varphi^\star_2}
		=
		\left( \frac{V_{,2}}{V} \right)^c
		\left( \frac{V}{V_{,2}} \right)^\star 
		\cos^2 \theta ,
	\end{eqnarray}
	where we have defined
	\begin{equation}
		\tan^2 \theta \equiv \frac{(V_{,2})^2}{(V_{,1})^2} .
	\end{equation}
	Eqs.~\eref{eq:1c1star}--\eref{eq:2c2star} can alternatively be derived
	without use of $C$ by comparing
	Eq.~\eref{e:NfromPhiPhiStar} with the
	formulas of Ref.~\cite{Gordon:2000hv}, which were derived using
	conventional perturbation theory.
	Applying~\eref{e:NfromPhiPhiStar}, we obtain
	\begin{equation}
		\label{e:Nphi}
		N_{,1}
		=
		\left( \frac{V}{V_{,1}} \right)^\star \cos^2 \theta ;
		\qquad
		N_{,2}
		=
		\left( \frac{V}{V_{,2}} \right)^\star \sin^2 \theta .
	\end{equation}
	To proceed, we require the second derivatives of $N$.
	These can be obtained directly from~\eref{e:Nphi},
	after use of Eqs.~\eref{eq:1c1star}--\eref{eq:2c2star}.
	We find
	\begin{eqnarray}
		\fl\nonumber
		\label{e:Nphiphi}
		N_{,11}
		=
		\left[ 1- \frac{V V_{,11}}{(V_{,1})^2} \right]^\star
		\cos^2 \theta 
		+ 2 \left( \frac{V}{V_{,1}} \right)^{\star 2}
		\cos^2 \theta \\ \hspace{-6mm}
		\mbox{} \times \left[
			\frac{V_{,11}}{V} \sin^2 \theta
			- \frac{V_{,1} V_{,12}}{V V_{,2}} \sin^4 \theta
			- \left(
				\frac{V_{,11}}{V} - \frac{V_{,22}}{V}
				+ \frac{V_{,2} V_{,12}}{V V_{,1}}
			\right)
			\cos^2 \theta \sin^2 \theta
		\right]^c .
	\end{eqnarray}
	An analogous expression for
	$N_{,22}$
	can be obtained after the simultaneous
	exchange $\{ 1 \leftrightarrow 2, \sin \leftrightarrow \cos \}$.
	The mixed derivative satisfies
	\begin{eqnarray}
		\fl\nonumber
		\label{e:Nphichi}
		N_{,12}
		=
		2 \left( \frac{V}{V_{,1}} \right)^\star
		\left( \frac{V}{V_{,2}} \right)^\star
		\cos^2 \theta \\ \nonumber\hspace{-6mm}
		\mbox{} \times
		\left[
			- \frac{V_{,11}}{V} \sin^2 \theta
			+ \frac{V_{,1} V_{,12}}{V V_{,2}} \sin^4 \theta
			+ \left(
				\frac{V_{,11}}{V} - \frac{V_{,22}}{V}
				+ \frac{V_{,2} V_{,12}}{V V_{,1}}
			\right)
			\cos^2 \theta \sin^2 \theta
		\right]^c \\ \hspace{-6mm}
		\mbox{} + \cos^2 \theta \left(
			\frac{V_{,2}}{V_{,1}} - \frac{V V_{,12}}{V_{,1}^2}
		\right)^c .
	\end{eqnarray}
	Now that the calculation is complete, we can drop the superscripts
	`$\star$' and `$c$,' since any background quantity is the same
	on either hypersurface. Once this is done
	it can be verified that (despite appearances)
	Eq.~\eref{e:Nphichi} is invariant under the exchange $1 \leftrightarrow
	2$.
		
	\section{Transport equations}
	\label{sec:transport}
	
	In this section we return to the problem of evolution between horizon
	exit and the time of observation, and supply the %missing
	prescription which connects the distribution of field values
	at these %disparate 
	two times.
	
	\subsection{Non-gaussian distribution in the single-field case}
	
	We begin by discussing the single-field system, which lacks the
	technical complexity of the two-field case, yet still exhibits
	certain interesting features which recur there.
	Among these features are
	the subtle difference between motion of the statistical mean
	and the background field value, and the hierarchy of moment evolution
	equations.	 Moreover, the structure of the moment mixing equations
	is similar to that which obtains in the two-field case.
	For this reason, the one-field scenario provides an instructive
	example of the techniques we wish to employ.
	
	Recall that we work in real space with a collection of
	comparably sized spacetime volumes,
	each with a slightly different expansion history, and the scatter in these 
	histories determines the microwave background anisotropy on 
	a given angular scale.
	Within each volume the smoothed background field $\varphi$ takes a
	uniform value described by a density function $P(\varphi)$,
	where in this section we are dropping the superscript `$\star$'
	denoting evaluation of spatially flat hypersurfaces.
	Our ultimate goal is to calculate the reduced bispectrum,
	$\fnl$, which describes the third moment of $P(\varphi)$.
	In the language of 
	probability this is the skewness,
	which we denote $\alpha$.
	A Gaussian distribution has skewness zero, and
	inflation usually predicts that the skew is small.
	For this reason, rather than seek a distribution
	with non-zero third moment,
	as proposed in Ref.~\cite{Fosalba:1999ha},
	we will introduce higher moments as
	perturbative corrections to the Gaussian. Such a procedure is
	known as a \emph{cumulant expansion}.

	The construction of cumulant expansions is a classical problem
	in probability theory.
	We seek a distribution	 with centroid
	$\varphi_0$, variance $\sigma^2$, and skew $\alpha$, with all higher
	moments
	determined by $\sigma$ and $\alpha$ alone.
	A distribution with suitable properties is
	\begin{equation}
		\label{e:P1D}
		P(\varphi)
		=
		P_g(\varphi)
		\left[
			1 + \frac{\alpha}{6 \sigma^3}
			H_3\left(\frac{\varphi-\varphi_0}{\sigma}\right)
		\right] ,
	\end{equation} 
	where
	\begin{equation}
		P_g(\varphi)
		\equiv
		\frac{1}{\sqrt{2\pi} \sigma} \exp
		\left[
			- \frac{\left(\varphi-\varphi_0\right)^2}{2 \sigma^2}
		\right]
	\end{equation}
	is a pure Gaussian and $H_n$ denotes the
	$n^\mathrm{th}$ Hermite polynomial, for which there are multiple
	normalization conventions.
	We choose to normalize so that
	\begin{equation}
		\label{eq:hermite-orthonormality}
		\int_{-\infty}^{\infty} \frac{1}{\sqrt{2\pi}} e^{-x^2/2} \,
		H_n(x) H_m(x) \; \d x =  n! \, \delta_{mn} ,
	\end{equation}
	which implies that the leading term of $H_n(x)$ is $x^n$.
	This is sometimes called the ``Probabilist's convention.''
	We define expectation values $\langle \cdots \rangle$ by
	the usual rule,
	\begin{equation}
		\langle F \rangle
		\equiv
		\int_{-\infty}^{\infty} P(\varphi) F \; \d x .
	\end{equation}
	The probability density function
	in Eq.~\eref{e:P1D} has the properties%
		\footnote{These formulas apply for arbitrary values of
		$\alpha$, and
		do not depend on the approximation that $\alpha$ is
		small. However, for large $\alpha$ the
		density function~\eref{e:P1D} may become negative for some values
		of $\varphi$. It then ceases to be a probability density in the
		strict sense. 
		This does not present a problem in practice, since
		we are interested in distributions which
		are approximately Gaussian, and for which $\alpha$ will
		typically be small. Moreover, our principal use of
		Eq.~\eref{e:P1D} is as a formal tool to extract evolution
		equations for each moment.
		For this reason we will not worry
		whether $P(\varphi)$ defines an honest probability density function
		in the strict mathematical sense.}
	\begin{equation}
		\label{e:P1Dnice}
		\langle 1 \rangle = 1 ,
		\quad
		\langle \varphi \rangle = \varphi_0 ,
		\quad \langle (\varphi-\varphi_0)^2 \rangle = \sigma^2 ,
		\quad \textrm{and}
		\quad \langle (\varphi-\varphi_0)^3 \rangle = \alpha .
	\end{equation}
	The moments $\varphi_0$, $\sigma$, and $\alpha$ may be time-dependent,
	so evolution of the probability density in
	time can be accommodated
	by finding evolution equations for these quantities.
	
	The density function given in Eq.~\eref{e:P1D}
	is well-known and has been applied in many situations.
	It is a solution to the problem of approximating a nearly-Gaussian
	distribution whose moments are known. 
	Eq.~(\ref{e:P1D}) is in fact the first two terms of the
	\emph{Gram--Charlier `A' series}, also sometimes called the
	\emph{Gauss--Hermite series}.%
		\footnote{In the physics literature, this series has sometimes
		erroneously been called the Edgeworth expansion.
		}
	In recent years it has found multiple applications to cosmology,
	of which our method is closest to that of Taylor \& Watts
	\cite{Taylor:2000ag}.
	Other applications are
	discussed in Refs.~\cite{Bouchet:1993xb,
	Bouchet:1995ez,Juszkiewicz:1993hm,Amendola:1996ny,Fosalba:1999ha,
	Taylor:2000ag,Matarrese:2000iz,Amendola:2001up,Watts:2002rz,
	LoVerde:2007ri,Seery:2006wk,Lam:2009nb}.
	For a review of the `A' series and related nearly-Gaussian probability
	distributions from an astrophysical perspective, see
	\cite{Blinnikov:1997jq}. In this paper, we will refer to
	Eq.~\eref{e:P1D} and its natural generalization to higher moments
	as the ``moment expansion.''
	
	In the slow-roll approximation,
	the field in each spacetime volume obeys a
	simple equation of motion
	\begin{equation}
		\label{e:1Dvelocity}
		\frac{\d \varphi}{\d N}
		=
		- \frac{\partial \ln V(\varphi)}{\partial \varphi}
		\equiv
		u(\varphi) ,
	\end{equation}
	where $N$ records the number of e-foldings of expansion.
	We refer to $u(\varphi)$ as the velocity field.
	Expanding $u$ about the instantaneous centroid $\varphi_0$ gives
	\begin{equation}
		u(\varphi) = u_0	+ u_{\varphi} (\varphi - \varphi_0)
			+ \frac{1}{2} u_{\varphi\varphi} (\varphi - \varphi_0)^2
			+ \cdots ,
	\end{equation}
	where
	\begin{equation}
		u_0 \equiv u |_{\varphi_0},
		\quad
		u_{\varphi} \equiv \left.\frac{\d u}{\d \varphi} \right|_{\varphi_0},
		\quad
		u_{\varphi\varphi} \equiv \left.\frac{\d^2 u}{\d \varphi^2}
			\right|_{\varphi_0} .
	\end{equation}
	The value of $\varphi_0$
	evolves with time,
	so each expansion coefficient is time-dependent.
	Hence, we do not assume that the velocity field is \emph{globally}
	well-described by a quadratic Taylor expansion, but merely that it
	is well-described as such in the neighborhood of the instantaneous
	centroid.	We
	expand the velocity field to second
	order, although in principle this expansion could be carried to
	arbitrary order.
	
	It remains to specify how the probability density evolves in time.
	Conservation of probability leads to the transport equation
	\begin{equation}
		\label{e:1DVlasov}
		\frac{\partial P}{\partial N}
		+ \frac{\partial}{\partial \varphi} ( u P )
		=
		0 .
	\end{equation}
	Eq.~\eref{e:1DVlasov} can also be understood as the limit
	of a Chapman--Kolmogorov process as the size
	of each hop goes to zero. % \cite{Taylor:2000ag}.
	It is well known---for example, from the study of
	Starobinsky's diffusion equation which forms the basis
	of the stochastic approach to inflation
	\cite{Starobinsky:1986fx}---that the choice of
	time variable in this equation is significant,
	with different choices corresponding to the selection of a
	temporal gauge.
	We have chosen to 
	use the e-folding time, $N$,
	which means that we are evolving the distribution on hypersurfaces
	of uniform expansion. These are the spatially flat hypersurfaces
	whose field perturbations enter the $\delta N$ formulas described in
	\S\ref{sec:computing_fnl}.
	
	In principle, Eq.~\eref{e:1DVlasov} can be solved
	directly. In practice it is simpler to extract equations
	for the moments of $P$, giving evolution equations for
	$\varphi_0$, $\sigma$ and $\alpha$. To achieve this,
	one need only resolve Eq.~\eref{e:1DVlasov}
	into a Hermite series of the form
	\begin{equation}
		\label{eq:hermite-resolved}
		P_g \sum_{n \geq 0} c_n H_n(\varphi-\varphi_0) = 0
	\end{equation}
	The Hermite polynomials are linearly independent,
	and application of the orthogonality
	condition~\eref{eq:hermite-orthonormality} shows that the $c_n$
	must all vanish. This leads to a hierarchy of equations
	$c_n = 0$, which we refer to as the moment hierarchy.
	At the top of the hierarchy, the equation $c_0 = 0$ is empty and
	expresses conservation of probability.
	
	The first non-trivial equation requires $c_1 = 0$ and yields
	an evolution equation for the centroid $\varphi_0$,
	\begin{equation}
		\frac{\d \varphi_0}{\d N}
		=
		u_0 + \frac{1}{2} u_{\varphi\varphi} \sigma^2 .
	\end{equation}
	The first term on the right-hand side drives the centroid along the
	velocity field,
	as one would anticipate
	based on the background equation of motion, Eq.~\eref{e:1Dvelocity}.
	However, the second term shows that the
	centroid is also influenced as the wings of the probability distribution
	probe the nearby velocity field. This influence is not
	captured by the background equation of motion.
	If we are in a situation with
	$u_{\varphi\varphi} > 0$, then the wings of the
	density function will be moving faster than the
	center. Hence, the velocity of the centroid will be
	larger than one might expect by restricting attention to
	$\varphi_0$.
	Accordingly, the
	mean fluctuation value is not following a solution to the background
	equations of motion.
	
	Evolution equations for the variance $\sigma^2$ and skew
	$\alpha$ are obtained after enforcing $c_2 = c_3 = 0$, yielding
	\begin{eqnarray}
		\label{e:EvoVar1D}
		\frac{\d \sigma^2}{\d N}
		=
		2 u_{\varphi} \sigma^2
		+ u_{\varphi\varphi} \alpha \\
		\label{e:EvoAlpha1D}
		\frac{\d \alpha}{\d N}
		=
		3 u_{\varphi} \alpha
		+ 3 u_{\varphi\varphi} \sigma^4
	\end{eqnarray}
	In both equations, the first term on the right-hand sides
	describes how $\sigma$ and $\alpha$ scale as the
	density function expands or contracts in response to the velocity
	field.
	These terms force $\sigma^2$
	and $\alpha$ to scale in proportion
	to the velocity field. Specifically, if we
	temporarily drop the second terms in each equation above, one finds that
	$\sigma^2 \sim u^2$ and $\alpha \sim u^3$. This precisely
	matches our expectation for the scaling of these quantities.
	Hence, these terms account for the Jacobians
	associated with
	infinitesimal transformations induced by the flow $u(\varphi)$. 
	
	For applications to inflationary non-Gaussianity, the second terms
	in~\eref{e:EvoVar1D} and~\eref{e:EvoAlpha1D} are more relevant.
	These terms describe how each moment is sourced by higher moments and
	the interaction of the density function with the velocity field.
	In the example above, if we are in a situation where
	$u_{\varphi\varphi} > 0$, the tails of the density function are
	moving faster than the core. This means that one tail is shrinking
	and the other is extending, skewing the probability density.
	The opposite occurs when $u_{\varphi\varphi} < 0$.
	These effects are measured by the second term in~\eref{e:EvoAlpha1D}.
	Hence, by expanding our PDF to the third moment, and our velocity
	field to quadratic order, we are able to construct a set of evolution
	equations which include the leading-order source terms for
	each moment.

	\subsection{The two-field case}
	\label{sec:two-field}

	There is little conceptually new as we move from one field to two.
	The new features are mostly technical in nature. Our primary
	challenge 
	is a generalization of the moment expansion to two
	fields, allowing for the possibility of correlation between
	the fields.	With this done, we can write
	down evolution equations whose structure is very similar to those found
	in the single-field case.

	The two-field system
	is described by a two-dimensional velocity
	field $u_i$, defined by
	\begin{equation}
		\label{e:DefU}
		u_i = \frac{\d \varphi_i}{\d N} ,
	\end{equation}
	where again
	we are using the number of e-folds $N$ as the time
	variable. The index $i$ takes values in $\{ 1,2 \}$.
	While we think it is likely that
	our equations
	generalize to any number of fields, we have only explicitly constructed
	them for a two-field system. As will become clear
	below, certain steps in this
	construction apply only for two fields, and hence we make no claims at
	present
	concerning examples with three or more fields.
	
	The two-dimensional transport equation is
	\begin{equation}
		\label{e:Vlasov}
		\frac{\partial P(\varphi_i,N)}{\partial N}
		+ \frac{\partial}{\partial \varphi_j}
		\left[ u_j  P(\varphi_i,N) \right]
		=
		0 .
	\end{equation}
	Here and in the following we
	have returned to our convention that
	repeated species indices
	are summed.
	As in the single-field case, we construct a probability distribution
	which is nearly Gaussian, but has a small non-zero skewness.
	That gives
	\begin{equation}
		P(\varphi_i,N) \equiv P_g(\varphi_i,N) P_{ng}(\varphi_i,N)
	\end{equation}
	where $P_g$ is a pure Gaussian distribution, defined by
	\begin{equation}
		\label{eq:gaussian-def}
		P_g(\varphi_i,N)
		=
		\frac{1}{2\pi \sqrt{\det \Sigma}}
		\exp \left[
			- \frac{1}{2} (\varphi_i - \Phi_i)
			(\Sigma^{-1})_{ij} (\varphi_j - \Phi_j)
		\right] .
	\end{equation}
	In this equation, $\Phi_i$ defines the center of the distribution
	and $\Sigma$ describes the covariance between the fields.
	We adopt a conventional parametrization in terms of
	variances $\sigma_i^2$ and a correlation coefficient $\rho$,
	\begin{equation}
		\label{eq:covariance-def}
		\Sigma \equiv \left( 
		\begin{array}{cc}
			\sigma_1^2 &	 \rho \sigma_1 \sigma_2 \\
			\rho \sigma_1 \sigma_2 & \sigma_2^2
		\end{array}
		\right) .
	\end{equation}
	The matrix $\sigma$ defines two-point correlations of the fields,
	\begin{equation}
		\langle (\varphi_i-\Phi_i) (\varphi_j-\Phi_j) \rangle
		= \Sigma_{ij} .
	\end{equation}
	
	All skewnesses are encoded in $P_{ng}$.
	Before defining this explicitly, it is helpful to pause and notice
	a complication inherent in
	Eqs.~\eref{eq:gaussian-def}--\eref{eq:covariance-def} which
	was not present in the single-field case.
	To extract a hierarchy of moment evolution equations from
	the transport equation, Eq.~\eref{e:1DVlasov},
	we made the expansion given in~\eref{eq:hermite-resolved} and argued that
	orthogonality of the Hermite polynomials implied the hierarchy
	$c_n = 0$. However, Hermite polynomials of the form
	$H_n[ (\varphi_i - \Phi_i)/\sigma ]$ are \emph{not} orthogonal
	under the Gaussian measure of Eq.~\eref{eq:gaussian-def}.
	Following an expansion analogous to Eq.~\eref{eq:hermite-resolved}
	the moment hierarchy would comprise linear combinations of the
	coefficients. The problem is essentially
	an algebraic question of Gram--Schmidt orthogonalization.
	
	To avoid this problem it is convenient to diagonalize the covariance
	matrix $\Sigma$, introducing new variables $X$ and $Y$
	for which Eq.~\eref{eq:gaussian-def} factorizes into the product
	of two measures under which the polynomials $H_n(X)$ and $H_n(Y)$
	are separately orthogonal. The necessary redefinitions are
	\begin{equation}
		\label{e:Xtrans}
		X
		\equiv
		\frac{1}{\sqrt{2(1+\rho)}}
		\left[
			\left( \frac{\varphi_1 - \Phi_1}{\sigma_1} \right)
			+ \left( \frac{\varphi_2 - \Phi_2}{\sigma_2} \right)
		\right]
	\end{equation}
	and
	\begin{equation}
		\label{e:Ytrans}
		Y
		\equiv
		\frac{1}{\sqrt{2(1-\rho)}}
		\left[
			\left( \frac{\varphi_1 - \Phi_1}{\sigma_1} \right)
			- \left( \frac{\varphi_2 - \Phi_2}{\sigma_2} \right)
		\right] .
	\end{equation}
	A simple expression for $P_g$ can be given
	in terms of $X$ and $Y$,
	\begin{equation}
		P_g
		=
		\frac{1}{2\pi} \exp
		\left(-\frac{X^2}{2}\right) \exp \left(-\frac{Y^2}{2}\right) .
	\end{equation}
	We now define the
	non-Gaussian factor, which encodes the skewnesses, to be
	\begin{eqnarray}
		\fl\nonumber
		P_{ng}
		\equiv
		1
		+ \frac{\alpha_{XXX}}{6} H_3(X) \\ \mbox{}
		+ \frac{\alpha_{XXY}}{2} H_2(X) H_1(Y)
		+ \frac{\alpha_{XYY}}{2} H_1(X) H_2(Y)
		+ \frac{\alpha_{YYY}}{6} H_3(Y) .
		\label{eq:nongaussian-def}
	\end{eqnarray}
	In these variables we find
	$\langle X^2 \rangle = \langle Y^2 \rangle = 1$,
	but $\langle X Y \rangle = 0$. In addition, we have
	\begin{equation}
		\fl
		\langle XXX \rangle = \alpha_{XXX} ,
		\quad
		\langle XXY \rangle = \alpha_{XXY} ,
		\quad
		\langle XYY \rangle = \alpha_{XYY} ,
		\quad \textrm{and} \quad
		\langle YYY \rangle = \alpha_{YYY} .
		\label{eq:XY-skewnesses}
	\end{equation}
	
	In order for Eq.~\eref{eq:nongaussian-def} to be useful,
	it is necessary to express the skewnesses associated with the
	physical variables $\varphi_i$ in terms of $X$ and $Y$.
	By definition, these satisfy
	\begin{equation}
		\langle (\varphi_i-\Phi_i) (\varphi_j-\Phi_j)
		(\varphi_k-\Phi_k)\rangle \equiv \alpha_{ijk} .
	\end{equation}
	After substituting for the definition of these quantities inside
	the expectation values in Eq.~\eref{eq:XY-skewnesses} we
	arrive at the relations
	\begin{eqnarray}
		\alpha_{XXX}
		=
		\frac{1}{2\sqrt{2}(1+\rho)^{3/2}}
		\left(
			\frac{\alpha_{111}}{\sigma_1^3}
			+ 3\frac{\alpha_{112}}{\sigma_1^2\sigma_2}
			+ 3\frac{\alpha_{122}}{\sigma_1\sigma_2^2}
			+ \frac{\alpha_{222}}{\sigma_2^3}
		\right) , \\
		\alpha_{XXY}
		=
		\frac{1}{2\sqrt{2(1-\rho)}(1+\rho)}
		\left(
			\frac{\alpha_{111}}{\sigma_1^3}
			+ \frac{\alpha_{112}}{\sigma_1^2\sigma_2}
			- \frac{\alpha_{122}}{\sigma_1\sigma_2^2}
			- \frac{\alpha_{222}}{\sigma_2^3}
		\right) , \\
		\alpha_{XYY}
		=
		\frac{1}{2\sqrt{2(1+\rho)}(1-\rho)}
		\left(
			\frac{\alpha_{111}}{\sigma_1^3}
			- \frac{\alpha_{112}}{\sigma_1^2\sigma_2}
			- \frac{\alpha_{122}}{\sigma_1\sigma_2^2}
			+ \frac{\alpha_{222}}{\sigma_2^3}
		\right) , \\
		\alpha_{YYY}
		=
		\frac{1}{2\sqrt{2}(1-\rho)^{3/2}}
		\left(
			\frac{\alpha_{111}}{\sigma_1^3}
			- 3\frac{\alpha_{112}}{\sigma_1^2\sigma_2}
			+ 3\frac{\alpha_{122}}{\sigma_1\sigma_2^2}
			- \frac{\alpha_{222}}{\sigma_2^3}
		\right) .
	\end{eqnarray}
	The moments $\Phi_i$, $\Sigma_{ij}$ and $\alpha_{ijk}$
	are time-dependent, but for clarity we will usually suppress this
	in our notation.
	
	Next we must extract the moment hierarchy, which governs evolution of
	$\Phi_i$, $\sigma_i$, $\rho$ and  $\alpha_{ijk}$.
	We expand the velocity field in a neighborhood of the
	instantaneous centroid $\Phi_i$ according to
	\begin{equation}
		\label{e:UTaylor}
		u_i(\varphi_j)
		=
		u_{i0}
		+ u_{ij} (\varphi_j-\Phi_j)
		+ \frac{1}{2} u_{ijk}	 (\varphi_j-\Phi_j) (\varphi_k-\Phi_k)
		+ \cdots ,
	\end{equation}
	where we have defined
	\begin{equation}
		u_{i0} \equiv u_i |_{\Phi_i} ,
		\quad
		u_{ij} \equiv \left. \frac{\partial u_i}{\partial \varphi_j}
		 	\right|_{\Phi_i} ,
		\quad \textrm{and} \quad
		u_{ijk} \equiv \left. \frac{\partial^2 u_i}
			{\partial \varphi_j \partial \varphi_k} \right|_{\Phi_i}
		.
		\label{e:UTaylor-definitions}
	\end{equation}
	As in the single-field case, these coefficients are functions of
	time and vary with the motion of the centroid. The expansion can be
	pursued to higher order if desired.
	
	Our construction of $X$ and $Y$ implies that
	the two-field transport equation can be arranged
	as a double Gauss--Hermite expansion,
	\begin{equation}
		\label{e:VlasovX}
		\frac{\partial P(\varphi_i,N)}{\partial N}
		+ \frac{\partial}{\partial \varphi_i}
			\left[ u_i \, P(\varphi_i,N) \right]
		=
		P_g \sum_{m,n \ge 0} c_{mn} H_m(X) H_n(Y)
		=
		0 .
	\end{equation}
	Because the Hermite polynomials are orthogonal in the measure
	defined by $P_g$, we deduce the moment hierarchy
	\begin{equation}
		c_{mn} = 0 .
	\end{equation}
	
	We define the ``rank" $r$ of each coefficient $c_{mn}$ by
	$r \equiv m+n$.
	We terminated the velocity field expansion at quadratic order, and
	our probability distribution included only the first three moments.
	It follows that only $c_{mn}$
	with rank five or less are nonzero.
	If we followed the velocity field to higher order, or included higher
	terms in the moment expansion,
	we would obtain non-trivial higher-rank coefficients.
	Inclusion of additional coefficients requires no qualitative modification
	of our analysis and can be
	incorporated in the scheme we describe below.
	
	A useful feature of the expansion in Eq.~\eref{e:VlasovX}
	is that the rank-$r$ coefficients give evolution equations for the
	order-$r$ moments.
	Written explicitly in components,
	the expressions that result from~\eref{e:VlasovX} are quite cumbersome.
	However, when written as field-space covariant expressions they
	can be expressed in a surprisingly compact form.
	
	\begin{description}
		\item[\textsf{Rank 0}]
		The rank-0 coefficient $c_{00}$ is identically zero.
		This expresses the fact that the total probability is conserved as
		the distribution evolves.
		
		\item[\textsf{Rank 1}]
		The rank-1 coefficients $c_{01}$ and $c_{10}$ give evolution
		equations for the centroid $\Phi_i$.	These equations can be
		written in the form
		\begin{equation}
			\label{e:Ev1}
			\frac{\d \Phi_i}{\d N}
			=
			u_{i0}
			+ \frac{1}{2} \Sigma_{jk} u_{ijk} .
		\end{equation}
		We remind the reader that here and below, terms like $u_{i0}$,
		$u_{ij}$ and $u_{ijk}$ represent the velocity field and its
		derivatives evaluated at the centroid $\Phi_i$.  The first term in
		\eref{e:Ev1} expresses the non-anomalous motion of the centroid, which
		coincides with the background velocity field
		of Eq.~\eref{e:DefU}.
		The second term describes how the wings of the probability
		distribution sample the velocity field at nearby points.	Narrow
		probability distributions have small components of $\Sigma$ and
		hence are only sensitive to the local value of $u_i(\varphi_j)$.
		Broad probability distributions have large components of $\Sigma$
		and are therefore more sensitive to the velocity field far from the
		centroid.

		\item[\textsf{Rank 2}]
		The rank-2 coefficients $c_{02}$, $c_{11}$ and $c_{20}$ give
		evolution equations for the variances $\sigma_i^2$ and the correlation
		$\rho$.  These can conveniently be packaged as evolution equations
		for the matrix $\Sigma$
		\begin{equation}
			\label{e:Ev2}
			\frac{\d \Sigma_{ij}}{\d N}
			=
			u_{ik} \Sigma_{kj}
			+ u_{jk} \Sigma_{ki}
			+ \frac{1}{2} \left(
				\alpha_{imn} u_{jmn} +	\alpha_{jmn} u_{imn}
			\right) .
		\end{equation}
		This equation describes the stretching and rotation of $\Sigma$ as
		it is transported by the velocity field.	
		It includes a sensitivity to the wings of the probability
		distribution, in a manner analogous to the similar term
		appearing in~\eref{e:EvoVar1D}. Hence the skew
		$\alpha_{ijk}$ acts as a source for the correlation matrix.
		
		\item[\textsf{Rank 3}]
		The rank-3 coefficients $c_{03}$, $c_{12}$, $c_{21}$ and $c_{30}$
		describe evolution of the moments $\alpha_{ijk}$.	 These are
		\begin{eqnarray}
			\nonumber
			\frac{\d \alpha_{ijk}}{\d N}
			=
			u_{in} \alpha_{njk}
			& + \Sigma_{jm} u_{imn} \Sigma_{nk} \\ & \mbox{}
			+ \textrm{cyclic permutations $i \rightarrow j \rightarrow k$} .
			\label{e:Ev3}
		\end{eqnarray}
		The first term describes how the moments flow into each other as
		the velocity field rotates and shears the $(X,Y)$ coordinate frame
		relative to the $\varphi_i$ coordinate frame. The second term
		describes sourcing of non-Gaussianity from inhomogeneities in the
		velocity field and the overall
		spread of the probability distribution.
	\end{description}
		
	Some higher-rank coefficients---in our case, those of ranks four and
	five---are also nonzero, but do not give any new evolution equations.
	These coefficients measure the ``error" introduced by truncating the
	moment expansion.	If we had included higher cumulants, these
	higher-rank coefficients would have given
	evolution equations for the higher moments of the probability
	distribution.	In general, all moments of the density function
	will mix so it is
	always necessary to terminate our expansion at a predetermined order---%
	both in cumulants and powers of the field fluctuation.
	The order we have chosen is sufficient to generate evolution equations
	containing both the leading-order behavior of the moments---namely, the
	first terms in Eqs.~\eref{e:Ev1}, \eref{e:Ev2} and~\eref{e:Ev3}---and
	the leading corrections, given by the latter terms in these equations.
		
	\section{Numerical results}
	\label{sec:numerics}

	At this point we put our new method into practice.
	We study two models
	for which the non-Gaussian signal is already known, using
	the standard $\delta N$
	formula.
	For each case we employ our method and compare it with results obtained
 	using
	$\delta N$. 
	To ensure a fair
	comparison, we solve
	numerically in both cases.
	Our new method employs the slow-roll approximation,
	as described above. Therefore,
	when using the $\delta N$ approach we produce
	results both with and without
	slow-roll simplifications.

	First consider double quadratic inflation, which was
	studied by Rigopoulos, Shellard \& van Tent
	\cite{Rigopoulos:2005xx,Rigopoulos:2005us}
	and later by Vernizzi \& Wands
	\cite{Vernizzi:2006ve}.
	The potential is
	\begin{equation}
		V(\phi, \chi)=\frac{1}{2} m_\phi^2 \phi^2
			+ \frac{1}{2} m_\chi^2 \chi^2 .
		\label{eq:double-quadratic}
	\end{equation}
	We use the initial conditions chosen in Ref.~\cite{Rigopoulos:2005xx},
	where $m_\phi/m_\chi = 9$,
	and the fiducial trajectory
	has coordinates
	$\phi^\star = 8.2 $ and $\chi^\star = 12.9$.

\begin{figure}
		\center{\includegraphics[width = 9.5cm]{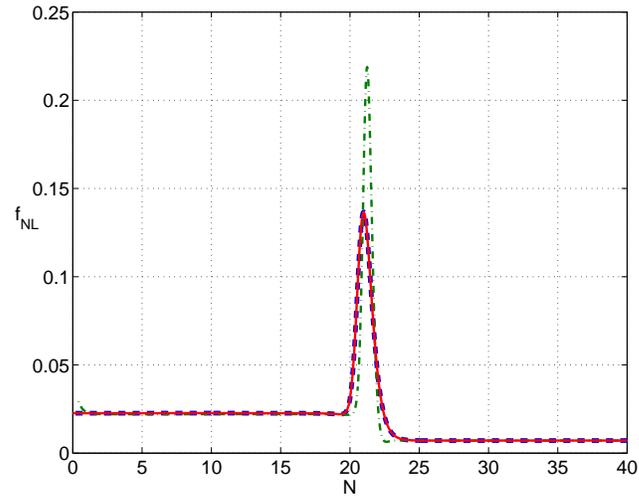}}
		\caption{Evolution of $\fnl$ in double quadratic inflation.
			The solid red line is obtained by
			numerically solving the moment transport equations
			obtained in \S\ref{sec:transport}.
			The blue dashed line and
			green dot-dashed line
			are the output of a
			numerical implementation
			of the standard $\delta N$ approach, with and without 
			slow roll respectively, using the fiducial picture.
			The red and blue lines lie on top of each other.
		\label{fig1}}
	\end{figure}

	We plot the evolution of $\fnl$
	in Fig.~\ref{fig1}, which also shows the prediction of the
	standard $\delta N$ formula (with and without employing slow roll
	simplifications).
	We implement the $\delta N$ algorithm using a finite
	difference method to calculate the derivatives of $N$.
	A similar technique was used in Ref.~\cite{Vernizzi:2006ve}.
	This model yields a very modest non-Gaussian signal, below unity
	even at its peak. If inflation ends away from the spike
	then $\fnl$ is practically negligible.

\begin{figure}
		\center{\includegraphics[width = 9.5cm]{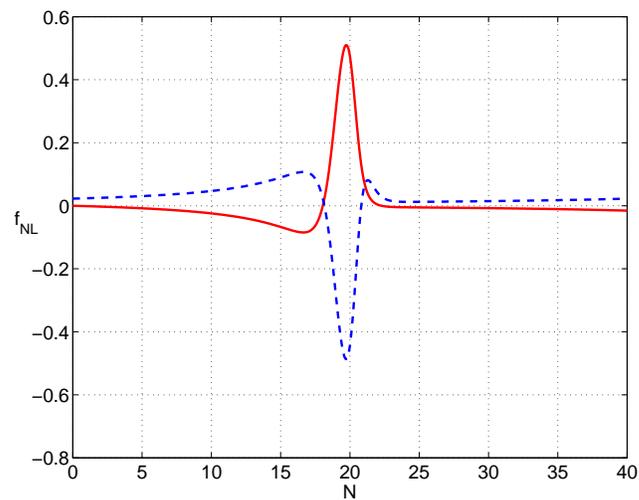}}
		\caption{Evolution of $\fnla$ (solid red line),
 			measuring the contribution of intrinsic non-linearities
			among the field fluctuations; and $\fnlb$
			(dashed blue line), measuring the contribution of the
			gauge transformation to $\zeta$.
		\label{fig2}}
	\end{figure}

	Eq.~\eref{e:zetaThree}
	shows that the method of moment transport allows us to separate
	contributions
	to $\fnl$ from the intrinsic non-Gaussianity of the field
	fluctuations, and non-linearities of the gauge transformation to
	$\zeta$.
	As explained in
	\S\ref{ss:sep_universe},
	we denote the former $\fnla$ and the latter $\fnlb$,
	and plot them separately in Fig.~\ref{fig2}.
	Inspection of this figure clearly shows that $\fnl$ is
	determined by a cancellation between two much larger components.
	Its final shape and magnitude are exquisitely sensitive to their
	relative phase.
	Initially,
	the magnitudes of $\fnla$ and $\fnlb$ grow, but their sum remains small.
	The peak in Fig.~\ref{fig1} arises from the peak of
	$\fnlb$, which is incompletely cancelled by $\fnla$.
	It is remarkable that $\fnla$ initially evolves in
	exact opposition to the
	gauge transformation, to which it is not obviously connected.

\begin{figure}
		\center{\includegraphics[width = 9.5cm]{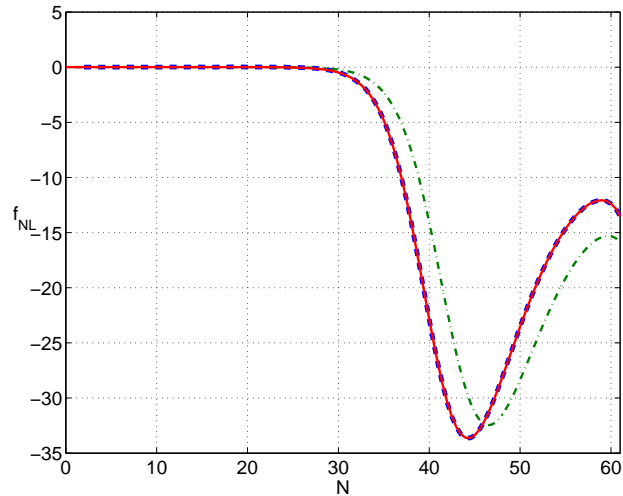}}
		\caption{Evolution of $\fnl$, for the potential
			$V=V_0 \chi^2 e^{-\lambda \phi^2}$
			(Example A from {\S}5 of Ref.~\cite{Byrnes:2008wi}).
			The solid red line represents the method of
			moment transport, whereas the
			blue dashed line and green dot-dashed line
			represents the
			output of conventional numerical $\delta N$ with and without
			slow-roll respectively.
			The red and blue lines are again coincident.
		\label{fig3}}
	\end{figure}

\begin{figure}
		\center{\includegraphics[width = 9.5cm]{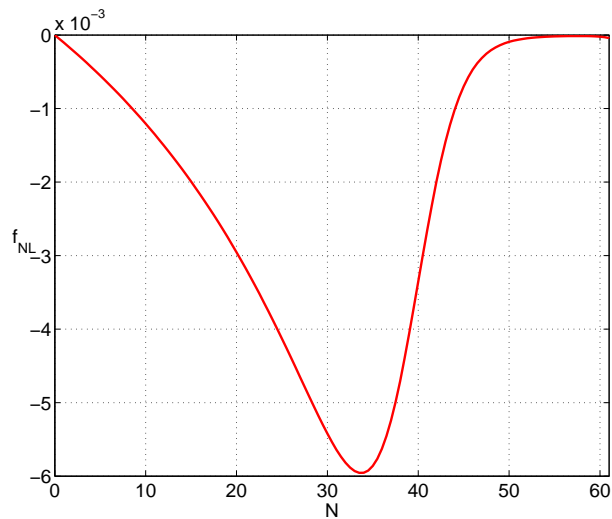}}
		\caption{Evolution of $\fnla$, for the potential $V = V_0 \chi^2
			e^{\lambda \phi^2}$.
			Comparison with Fig.~\ref{fig3} shows that $\fnl$
			is totally dominated by $\fnlb$.
		\label{fig4}}
	\end{figure}

	In the double quadratic model, $\fnl$ is always small. However,
	it has recently been shown by Byrnes {\etal} that a large
	non-Gaussian signal can be generated
	even
	when slow-roll is a good
	approximation
	\cite{Byrnes:2008wi,Byrnes:2008zy}.
	The conditions for this to occur are incompletely understood, but
	apparently require a specific
	choice of potential and strong tuning of
	initial conditions.
	In Figs.~\ref{fig3}--\ref{fig4}
	we show the evolution of $\fnl$ in a model with the potential
	\begin{equation}
		V = V_0 \chi^2 e^{-\lambda \phi^2} ,
	\end{equation}
	which corresponds to Example A of
	Ref.~\cite[{\S}5]{Byrnes:2008wi} when we choose
	$\lambda=0.05$ and initial conditions
	$\chi^\star=16$, $\phi^\star=0.001$.
	It is clear that the agreement is exact.		
	In this model, $\fnl$ is
	overwhelmingly dominated by the contribution from the
	second-order gauge transformation, $\fnlb$, as shown
	in Fig.~\ref{fig4}.
	This conclusion applies equally to the other large-$\fnl$ examples
	discussed in Refs.~\cite{Byrnes:2008wi,Byrnes:2008zy},
	although we make no claim that this is a general phenomenon.
		
	In conclusion, Figs.~\ref{fig1} and~\ref{fig3} show
	excellent
	agreement between our new
	method and the outcome of the numerical $\delta N$ formula.
	These figures
	also compare the moment transport
	method and $\delta N$
	without the slow-roll approximation.
	We conclude that the slow-roll estimate remains broadly accurate
	throughout the entire evolution.

	\section{Discussion}
	\label{s:conclusions}
	
	Non-linearities are now routinely extracted from all-sky observations
	of the microwave background anisotropy. Our purpose in this paper has
	been to propose a new technique with which to predict the observable
	signal. Present data already give interesting
	constraints on the skewness parameter $\fnl$, and over the next several
	years we expect that the \emph{Planck} survey satellite will
	make these constraints very stringent. It is even possible that
	higher-order moments, such as the kurtosis parameter $\gnl$
	\cite{Sasaki:2006kq}
	will become better constrained
	\cite{Desjacques:2009jb}.
	To meet the need of the observational community
	for comparison with
	theory, reliable estimates of these non-linear quantities will be
	necessary for various models of early-universe physics.

	A survey of the literature suggests that the
	`conventional' $\delta N$ method,
	originally introduced by Lyth \& Rodr\'{\i}guez,
	remains the method of choice for analytical study of
	non-Gaussianity.
	In comparison, our proposed moment transport method exhibits
	several clear differences.
	First,
	the conventional method functions best when we base the $\delta N$
	expansion on a flat hypersurface immediately after horizon exit.
	In our method, we make the opposite choice and
	move the flat hypersurface as close as possible
	to the time of observation.
	After this, the role of the $\delta N$ formula is
	to provide no more than the non-linear gauge transformation between
	field fluctuations and the curvature perturbation.
	We substitute the method of moment transport to evolve
	the distribution of field fluctuations between horizon exit and
	observation. 
%	We believe this carries extra information which is
%	neglected by the conventional $\delta N$ formula.
%	In particular, the transport equation is sensitive to an
%	approximate distribution of spatial gradients because it is aware
%	that nearby spacetime volumes may find themselves located at
%	different values of the field.%
%		\footnote{We emphasize, however, that there is no reason to
%		suppose the distribution function will respond in the same way
%		as it would, had gradient terms been retained in the equation of
%		motion.}
%	The terms discussed in \S\ref{sec:two-field} which correct
%	the background motion of $\Phi_i$, $\Sigma_{ij}$ and
%	$\alpha_{ijk}$ can be thought of as a response to the distribution of
%	spatial gradients in a typical inflating volume.
%	This information is discarded in the simplest $\delta N$ formula.

	Second, in integrating the transport equation one uses
	an expansion of the velocity field such as the one given in
	Eqs.~\eref{e:UTaylor}--\eref{e:UTaylor-definitions}.
	This expansion is refreshed at each step of integration,
	so the result is related to conventional perturbative calculations
	in a very similar way to renormalization-group improved perturbation
	theory
	\cite{GellMann:1954fq}.
	In this interpretation, derivatives of $u_i$ play the role of
	couplings. At a given order, $m$, in the moment hierarchy,
	the equations for lower-order moments function as renormalization group
	equations for the couplings at level-$m$, resumming
	potentially large terms before they spoil perturbation theory.
	This property is shared with any formalism such as $\delta N$
	which is non-perturbative in time evolution,
	but may be an advantage in comparison with perturbative methods.
	We also note that
	although $\delta N$ is non-perturbative as a point of principle,
	practical implementations are frequently perturbative.
	For example, the method of Vernizzi \& Wands
	\cite{Vernizzi:2006ve}
	and Battefeld \& Easther
	\cite{Battefeld:2006sz}
	depends on the existence of quantities which are conserved only
	to leading order in $\epsilon N$, and can lose accuracy after
	$N \sim \epsilon^{-1}$ e-foldings.

	Numerical calculations confirm that our method gives results
	in
	excellent
	agreement with existing techniques.	
	As a by-product of our analysis, we note that the large non-gaussianities
	which have recently been observed in sum- and product-separable
	potentials
	\cite{Byrnes:2008wi,Byrnes:2008zy}
	are dominated by non-linearities from the second-order part of the
	gauge transformation from $\delta\varphi_i$ to $\zeta$.
	The contribution from intrinsic non-linearities of the field
	fluctuations, measured by the skewnesses $\alpha_{ijk}$,
	is negligible.
	In such cases one can obtain a useful formula for $\fnl$
	by approximating the field distribution as an exact Gaussian.
	The non-Gaussianity produced in such cases arises from a
	distortion of comoving hypersurfaces with respect to adjacent
	spatially flat hypersurfaces.
	
	Our new method joins many well-established techniques for estimating
	non-Gaussian properties of the curvature perturbation. In our experience,
	these techniques give comparable estimates of $\fnl$, but they do not
	exactly agree. Each method invokes different assumptions, such as the
	neglect of gradients or the degree to which time dependence can be
	accommodated. The mutual scatter between different methods can be attributed
	to the theory error inherent in any estimate of $\fnl$.  The comparison
	presented in \S\ref{sec:numerics} shows that while all of these
	methods slightly disagree, the moment transport method gives good agreement
	with other established methods.

	\ack
	DM is supported by the Cambridge Centre for Theoretical Cosmology
	(CTC).
	DS is funded by STFC.
	DW acknowledges support from the CTC.
	We would like to thank Chris Byrnes, Jim Lidsey and Karim Malik
	for helpful conversations.

	\section*{References}

\providecommand{\href}[2]{#2}\begingroup\raggedright\endgroup
	
\end{document}